\documentclass[journal,onecolumn]{IEEEtran}
\usepackage{graphics}
\usepackage{graphicx}
\usepackage{url}
\usepackage{setspace}

\title{Optimal Gaussian Filter for Effective Noise Filtering}
\author{Sunil Kopparapu and M Satish
\thanks{Sunil Kumar Kopparapu and M Satish are with the TCS Innovation Labs - Mumbai, Yantra Park, Thane (West), Maharastra, INDIA.
Email: SunilKumar.Kopparapu@TCS.Com}}

\newcommand{\fs}{f_{s}}
\newcommand{\fmax}{f_{max}}
\newcommand{\noise}{n}
\newcommand{\Noise}{N}
\newcommand{\Signal}{X}
\newcommand{\signal}{x}
\newcommand{\x}{\signal}
\newcommand{\g}{f}
\renewcommand{\S}{{\cal S}}
\newcommand{\BW}{{\cal B}}
\newcommand{\slength}{{\cal N}}
\newcommand{\flength}{{\cal M}}

\renewcommand{\a}{\alpha}
\newcommand{\dt}{m}
\newtheorem{mynote}{Note}

\begin{document}
\maketitle

\doublespace
\begin{abstract}

In this paper we show that the knowledge of noise statistics 
contaminating a signal can be effectively used to choose an optimal
 Gaussian filter to 
eliminate noise. Very specifically, we show that the additive white 
Gaussian noise (AWGN) contaminating a signal can be filtered best by 
using a Gaussian filter of specific characteristics. The design of the 
Gaussian filter bears relationship with the noise 
statistics and also some basic information about the signal. We first 
derive a relationship between the properties of the Gaussian filter, 
noise statistics and the signal and later show through experiments that 
this relationship can be used effectively to identify the optimal 
Gaussian filter that can effectively filter noise.
\end{abstract}


\section{Introduction}

Signal smoothing or noise filtering or denoising 
has been an area of active research and continues to hold 
the attention of researchers in various fields, for example,
\cite{Crisan_Kouritzin_Xiong_2008,Oktem_Egiazarian_Lukin_Ponomarenko_Tsymbal_2007,Buades_Silva_Santos_2010,Huang_Wang_Long_2009,Yang_Wei_2010}.
Noise is inherent in signals 
\cite{Bruni_Piccoli_Vitulano_2008,Narayana:2009:ENM:1946497.1946503}
and a necessary first step is noise removal before any other processing can
take place. A successful pre-processing step to remove noise improves the
performance of the {\em actual processing} on the signal \cite{lajish_2010}.
There are essentially two ways of taking
care of noise in the signal, namely, (a) pre-processing of the signal to
enable noise removal or (b) use of a set of robust algorithms that can
compensate for the inherent noise. In signal processing 
literature pre-processing of the signal is
the preferred approach.

\subsection{Problem}
\label{sec:problem}
Let $\Signal = [ \signal_1, \signal_2, \cdots, \signal_N ]$ be a band 
limited ($\BW$) 
digitized signal which is sampled at a sampling frequency of $\fs$ and
Let 
$\Noise = [ \noise_1, \noise_2, \cdots, \noise_N ]$ be the noise sequence.
Further assume that $\{\noise_{i} \}_{i=1}^{N}$ is 
Gaussian distributed with mean $\mu_\Noise$ and variance 
$\sigma^2_\Noise$. Let 
 \begin{equation}\Signal_\Noise = \Signal + 
\Noise \label{eq:tdomain}
 \end{equation} 
represent the signal $\Signal$ 
contaminated by AWGN $\Noise$. 
Now the problem can be stated as, given 
$\Signal_\Noise$ estimate $\hat{\Signal}$ such that the error in the 
estimate is minimum, namely 
 \begin{equation} minarg_{\hat{\Signal}} || 
\Signal - \hat{\Signal} ||^2
 \end{equation} 
Typically the process of 
estimating $\hat{\Signal}$ given the noise contaminated 
${\Signal_\Noise}$ is called noise filtering or denoising. 
We will restrict our 
discussion, in this paper to the usage of a Gaussian smoothing filter for noise 
removal.
We 
describe Gaussian 
filtering in Section \ref{sec:gaussian_filtering} which is characterized 
by $\sigma_\g$ which determines the amount of smoothing. We build theory 
in Section \ref{sec:our_approach} 
which allows identification of an optimal $\sigma_\g^{opt}$. We show 
experimentally how the identification of the actual Gaussian filter can 
be found in Section \ref{sec:experimental_results} and conclude in 
Section \ref{sec:conclusions}.

\section{Gaussian Smoothing}
\label{sec:gaussian_filtering}

A Gaussian filter is parametrized by its means $\mu_f$ and variance 
$\sigma_f^2$ and represented by
 \begin{equation}
G_f(\mu_f, \sigma_f^2, t) = \frac{1}{\sqrt{2\pi\sigma_f^2}}
\exp^ { - \left \{ \frac{ (t-\mu_f)^2}{2 \sigma_f^2} \right \} }
\label{eq:gaussian_filter}
 \end{equation}
\begin{mynote}
Given $\mu_f$ and $\sigma_f^2$ one can construct a Gaussian filter 
(\ref{eq:gaussian_filter}) with $t$ running between $[-\infty, \infty]$.  
\end{mynote}
\begin{mynote}
It 
is well known that spanning $t$ between $-3 \sigma_f$ and $3 \sigma_f$ 
covers $99.7$ \% of the total area under the Gaussian. 
\end{mynote}
So we can approximate 
$ G_f(\mu_f, \sigma_f^2, t)$ from ${t=-\infty}$ to ${\infty}$ 
as
$G_f(\mu_f, \sigma_f^2, t)$ from ${t=-3 \sigma_f}$ to ${3 \sigma_f}$ 
for the purpose of 
discussion and subsequent experimentation. Let the discrete version of 
$G_f(\mu_f, \sigma_f^2, t)$ from ${t=-3 \sigma_f}$ to ${3 \sigma_f}$ 
be 
represented by 
$G_f[\mu_f, \sigma_f^2,$ $\dt]$ from 
$
\dt = {-\lceil3 \sigma_f\rceil}
$ to 
$
{\lceil 3  \sigma_f \rceil}$, where $\lceil \bullet \rceil$ represents the ceil
of $\bullet$.
Let 
$\Signal_{\Noise}$ 
smoothed with $G_f[\mu_f, \sigma_f^2, \cdot]$ 
result in $\hat{\Signal}^{\sigma_f^2}$, 
namely, 
 \begin{eqnarray}
 \hat{\Signal}_{{k}}^{\sigma_f^2} &= 
&\sum_{i = -\lceil 3 \sigma_f^2 \rceil + k}^{\lceil 3 \sigma_f^2 \rceil +k} \Signal_{\Noise_{i}} 
G_f[\mu_f, \sigma_f^2, i-k] \nonumber \\ 
& = & \sum_{i = - \lceil3 \sigma_f^2 \rceil
+k}^{3 \lceil \sigma_f^2 \rceil +k} {(\x_i + \noise_{i})} G_f[\mu_f,
\sigma_f^2, i-k] \nonumber \\
\label{eq:filtered}
 \end{eqnarray} 
 for $k = 1, 2, \cdots, N$. Let the error 
in the estimate be 
\begin{equation} 
E_{\sigma^2_f} = \frac{1}{N} \sum_{k=1}^N \left (\Signal_{k} - 
\hat{\Signal}_{k}^{\sigma_f^2} \right )^2 \label{eq:error} 
\end{equation}

We hypothesize that one can achieve an optimal estimate 
$\hat{\Signal}_{k}^{\sigma^2_f}$ for some $\sigma^2_f$ 
such that $E_{\sigma^2_f}$ is minimized. 
We further hypothesize that
$\sigma_f^2$ is based on the 
variance of the noise affecting the signal and some properties of the signal. 
Specifically, $\sigma_f^2$ is 
dependent directly or indirectly on $\sigma_\Noise^2$ and $\BW$.  

\section{Our Approach}
\label{sec:our_approach}

In the frequency domain we can write (\ref{eq:tdomain}) as
 \begin{equation}
\Signal_\Noise(\omega)=\Signal(\omega)+\Noise(\omega)
 \end{equation}
 and the Gaussian filter as

 \begin{equation}
G(\omega)=\exp{\left (\frac{-\omega^2 \sigma_\g^2}{2} \right )}
\label{eq:gaussian_frequency}
 \end{equation} The estimate of the signal 
$\hat{\Signal}_\Noise(\omega)$ due to filtering by Gaussian filter can be
written as
 \begin{equation}
\hat{\Signal}_\Noise(\omega)=\Signal(\omega) G(\omega) + \Noise(\omega) G(\omega)
 \end{equation} The error in the filtered output is given by
 \begin{eqnarray}
E(\omega) &=& \Signal(\omega) - \hat{\Signal}_\Noise(\omega)   \nonumber \\
& = & \underbrace{\Signal(\omega)\left[1-G(\omega)\right]}_{\mbox{Signal
Distortion}} + \underbrace{N(\omega)G(\omega)}_{\mbox{Noise Smoothing}}
\label{eq:error_components}
 \end{eqnarray}
 As seen in (\ref{eq:error_components}) the error in the estimate ($E(\omega)$)
due to 
filtering has two components namely, one due to distortion of signal 
($\Signal(\omega)\left[1-G(\omega)\right]$)
and 
the other due to the reminiscent noise  ($N(\omega)G(\omega)$)
in the signal after filtering. 
Let $P_{\bullet}$ denote the power in the signal $\bullet$, then
input and output signal to noise ($\S$) ratios are given by 
 \begin{eqnarray} 
\S_i & = & \frac{P_\Signal}{P_{\Noise}} \nonumber \\ 
\S_o &= & \frac{P_\Signal}{P_{\Signal} - P_{\hat{\Signal}}} = 
\frac{P_\Signal}{P_{E}}
 \end{eqnarray}
\begin{mynote}
For a certain $\sigma_\g^2$,  the Gaussian filter is able to filter the signal
such that $\S_o > \S_i$. Namely,
simultaneously 
remove the
noise and 
not distort the signal. 
\end{mynote}
\begin{mynote}
If we increase $\sigma_\g^2$ then the cutoff frequency and the 
bandwidth of Gaussian filter will decrease as seen in 
(\ref{eq:gaussian_frequency}) and subsequently this will lead to more 
noise removal but on same account the signal distortion will also increase. 
\end{mynote}
In the limiting case when 
$\sigma_\g \rightarrow 0$,  we have an all pass filter and hence 
$\S_o = \S_i$. 
Let for some $\sigma_\g^2 = \sigma_{\g R}^2$ 
$\S_o = \S_i$, such that  
if we increase $\sigma_\g^2$ further then $\S_o < \S_i$. 
One can hypothesize that for $\sigma_\g^2$ in the range 
$[0,\sigma_{\g R}^2]$, $\S_o > \S_i$. 
We further hypothesize that there exists a 
$\sigma_{\g,{opt}}^2$ (in the range $[0,\sigma_{\g R}^2]$) for which $\S_o$ 
peaks to achieve $\S_{o}^{max}$. We show through curve fitting and later 
experimentally that we can determine the optimal $\sigma_{\g,opt}^2$ 
such that $\S_o$ is maximized.

\subsection{Determining $\sigma_{\g,opt}^2$}

With an aim to identify $\sigma_{\g,opt}^2$ the optimal choice of 
Gaussian filter to remove noise we constructed three different signals 
($\Signal$) with different bandwidths ($\BW$). We constructed the noisy 
signal ($\Signal_\Noise$) by appending $\Signal$ with $\Noise$ with 
varying $\sigma_\Noise^2$. For each of this noisy signal we used 
different $\sigma_\g^2$ Gaussian to filter noise and for each of this 
$\S_o$ is computed.
The band limited $\Signal$  is constructed by first generating 
a random sequence of length $\slength$ having a normal distribution with mean
zero and variance one. This random signal is smoothened using a filter of length $\flength
(<< \slength)$. The impulse response of the smoothing filter is given by
\begin{eqnarray}
h(\dt) &=& {1} \;\; \mbox{for}\;\; 0 \le \dt \le \flength-1 \nonumber \\
&=& 0 \;\; \mbox{otherwise}
\end{eqnarray}
Note that if we take a $\slength$ point DFT of this smoothed signal, then most of the energy is
limited to $\fs/\flength$ Hz or $\slength/\flength$ points. We cut off the high frequency region of the signal, namely,
we set the points from  $\slength /\flength$ to $(\slength-\slength /\flength)$
to zero. The inverse DFT of this low-pass filtered signal is the test signal with maximum frequency $\fmax =$
$\fs/\flength$ Hz. Note that different values of $\flength$ produce a filtered signal with different $\fmax$ and 
hence bandwidths ($\BW$).
In this manner we constructed three different signals, each  of length 
$\slength = 1024$ with $\flength =5, 7, 10$. We denote these three signals as
$\Signal_{5}, \Signal_{7}$ and $\Signal_{10}$ having $\fmax$ of 
$\frac{\fs}{5}$,
$\frac{\fs}{7}$,
$\frac{\fs}{10}$ Hz
respectively. 
An additive white Gaussian 
noise with $\sigma_\Noise^2 = 30, 35$ and $40$ denoted by $\Noise_{30}, 
\Noise_{35}, \Noise_{40}$ is generated.
In all we had $9$ $\Signal_\Noise$ as our test bed. Namely,
 $\Signal_\Noise^1 = \Signal_5 + \Noise_{30}$, 
 $\Signal_\Noise^2 = \Signal_5 + \Noise_{35}$, 
 $\Signal_\Noise^3 = \Signal_5 + \Noise_{40}$, 
 $\Signal_\Noise^4 = \Signal_7 + \Noise_{30}$, 
 $\Signal_\Noise^5 = \Signal_7 + \Noise_{35}$, 
 $\Signal_\Noise^6 = \Signal_7 + \Noise_{40}$, 
 $\Signal_\Noise^7 = \Signal_{10} + \Noise_{30}$, 
 $\Signal_\Noise^8 = \Signal_{10} + \Noise_{35}$, 
 $\Signal_\Noise^9 = \Signal_{10} + \Noise_{40}$.

\begin{figure} 
 \centerline{\includegraphics[width=0.5\textwidth]{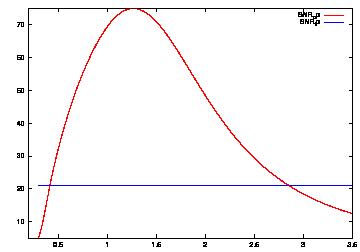}} 
 \caption{The $\S_o$ of filtered $\Signal_\Noise^2$ for different values 
of $\sigma_\g^2$.}
 \label{fig:snr_plot}
 \end{figure}

 These signals $\{ \Signal_\Noise^k \}_{k=1}^9$ are denoised using a 
Gaussian filter (\ref{eq:gaussian_filter}) with different $\sigma_\g^2$. 
We varied $\sigma_\g^2$ from $0.3$ to $3.5$ in steps 
of $0.01$ ($320$ data points). For all 
these filtered output signal, namely, $\hat{X}$, the $\S_o$ is 
calculated. Fig. \ref{fig:snr_plot} shows the $\S_o$ of the filtered 
$\Signal_\Noise^2$ for different values of $\sigma_\g^2$. The x-axis shows 
the different values of $\sigma_\g^2$ and the bell shaped curve is the 
$\S_o$; also $\S_i$ ($23$ dB) is shown as a horizontal line. 
We had $320$ $\S_o$ for varying $\sigma_\g^2$ for each of the $9$ 
noisy signals. We now try to fit a curve so as to relate the $\S_o$ in terms of 
$\BW$, $\S_i$ and 
$\sigma_\g^2$.  
We did this in two steps using \cite{web:curve_fit}.

\begin{enumerate}
 \item[Step 1] For a fixed $\BW$,  we fit a 3-D curve to relate $\S_o$, 
$\S_i$ 
and $\sigma_\g^2$ for  $\BW = 5, 7, 10$ separately
using the reciprocal full quadratic function\footnote{Experimented with
several functions before converging onto the reciprocal full quadratic function}, namely, 
\begin{eqnarray}
\S_o &= & \left \{ a_\BW+b_\BW \sigma_\g + c_\BW \S_i+ d_\BW {\sigma_\g}^2
\right .  \nonumber \\
&& \left . +f_\BW {\S_i}^2+g_\BW \sigma_\g \S_i \right \}^{-1}
\label{eq:cf1}
 \end{eqnarray}
with minimize
the sum of squared absolute error criteria. 
For each $\BW = 5, 7, 10$ we obtained a set of coefficients 
$a, b, c, d, f$ and $g$, so in all we had $18$ coefficients, namely, 
$A = [a_5, a_7, a_{10}]$,
$B = [b_5, b_7, b_{10}]$,
$C = [c_5, c_7, c_{10}]$,
$D = [d_5, d_7, d_{10}]$,
$F = [f_5, f_7, f_{10}]$ and
$G = [g_5, g_7, g_{10}]$

\item [Step 2] We then fit a quadratic curve 
for each coefficient set, namely, $A, B, C, D, F, G$  and $\BW$ separately.
Using $A$ we found that $a$ in (\ref{eq:cf1}) is related to $\BW$ as 
$a=\a_1 + \a_2 \BW+ \a_3{\BW}^2$. 
Similarly coefficients
$b, c, d, f, g$ can be written in terms of $\BW$. Namely, 
 \end{enumerate}
\begin{eqnarray}
a_\BW &=& (0.8364  -1.504 \BW + 4.017 \BW^2) \times {10}^{-1} \nonumber \\
b_\BW &=& (-0.1790  -2.572 \BW -4.164 \BW^2) \times {10}^{-1} \nonumber \\
c_\BW &=& (-0.4596 + 3.313 \BW -7.653 \BW^2)\times {10}^{-2} \nonumber \\
d_\BW &=& (0.7983 - 8.658 \BW + 1.575  \BW^2) \times {10}^{-2} \nonumber \\
f_\BW &=& (0.7481 -6.817 \BW + 17.04 \BW^2)\times {10}^{-4} \nonumber \\
g_\BW &=& (0.5562 - 2.510 \BW +  6.352 \BW^2) \times {10}^{-3} \nonumber \\
\label{eq:cf2}
\end{eqnarray}

 Now we have (\ref{eq:cf1}), we get $\sigma_{\g,opt}$ by 
differentiating (\ref{eq:cf1}) with
respect to
$\sigma_\g$ and setting \[ \frac{\partial \S_o}{\partial\sigma_\g} = 0 \]
namely, 
\begin{equation}
\sigma_{\g,opt}=-\frac{g_\BW \S_i+b_\BW}{2d_\BW}
\label{eq:sopt}
\end{equation}
where $g_\BW$, $b_\BW$, $d_\BW$ are given in (\ref{eq:cf2}). We get
$\S_o^{max}$ by substituting the value of $\sigma_{\g,opt}$ in
(\ref{eq:cf1}), namely,
\begin{eqnarray}
\S_o^{max}&=&\left \{ a_\BW +b_\BW \sigma_{\g,opt} + c_\BW \S_i+ d_\BW 
{\sigma_{\g,opt}}^2
\right .
\nonumber \\ 
&&\left . 
+f_\BW {\S_i}^2+g_\BW \sigma_{\g,opt} \S_i \right \}^{-1}
\label{eq:ssnr}
\end{eqnarray}

\section{Experimental Results}
\label{sec:experimental_results}

We conducted a number of experiments to verify the correctness of
(\ref{eq:sopt}) and (\ref{eq:ssnr}) in identifying $\sigma_{\g,opt}$ and
$\S_o^{max}$ respectively, these results are shown 
in Table \ref{tab:table1} and Table \ref{tab:table2}.
\begin{table}
\begin{center}
\begin{tabular}{|c|c|c|c|c|c|c|}
\hline \hline
$\BW$ & $\sigma_\Noise$ & $\S_i$ & $\sigma_{\g,opt}$  &  $\sigma_{\g,opt}$& 
$\S_{o}^{max}$ & $\S_{o}^{max}$ \\
&  &  & (\ref{eq:sopt}) &  & (\ref{eq:ssnr})&  \\
\hline \hline
$10$ & $30$ & $28.6$ & $1.23$ & $1.18$ & $96.2$ & $95.5$ \\ \hline
$10$ & $35$ & $21.0$ & $1.33$ & $1.26$ & $74.5$ & $75.0$ \\ \hline
$10$ & $40$ & $16.1$ & $1.39$ & $1.34$ & $59.8$ & $60.9$ \\ \hline \hline
$7$ & $30$ & $39.2$ & $0.90$ & $0.87$ & $95.8$ & $96.3$ \\ \hline
$7$ & $35$ & $28.8$ & $0.97$ & $0.92$ & $74.3$ & $75.6$ \\ \hline
$7$ & $40$ & $22.0$ & $1.00$ & $0.98$ & $59.9$ & $61.3$ \\ \hline \hline
$5$ & $30$ & $52.2$ & $0.65$ & $0.65$ & $90.2$ & $91.4$ \\ \hline
$5$ & $35$ & $38.4$ & $0.69$ & $0.69$ & $70.5$ & $72.2$ \\ \hline
$5$ & $40$ & $29.4$ & $0.72$ & $0.72$ & $57.2$ & $58.9$ \\ \hline
\end{tabular}
\caption{Comparison of actual $\sigma_\g$, $\S_o^{max}$ with derived
$\sigma_\g$ using (\ref{eq:sopt}), $\S_o^{max}$ using (\ref{eq:ssnr}).}
\label{tab:table1}
\end{center}
\end{table}
Table \ref{tab:table1} tries to access the goodness of the curve fit, namely,
the choice of the curve and the construction of (\ref{eq:sopt}) 
and (\ref{eq:ssnr}) from the data. 
As can be seen, the column four ($\sigma_{\g,opt}$ calculated from
(\ref{eq:sopt})) and column five (actual
$\sigma_{\g,opt}$ computed from the data) are very close to each other. This is
to be expected when the choice of the curve to fit the data is good. 
However to verify the validity of our approach to identify the $\sigma_{\g,opt}$
we conducted another set of experiments. We generated several 
test
signals with different $\flength$ and
$\Noise$ with different $\sigma_\Noise^2$, such that these test 
signals were not part of the signals used to
construct (\ref{eq:sopt}) using curve fitting.  As can be seen in Table
\ref{tab:table2}, the estimation of $\sigma_{\g,opt}$ using (\ref{eq:sopt}) is very
close to the actual $\sigma_{\g,opt}$ for all signals in Table \ref{tab:table2}. 
As expected, a similar match is seen for $\S_o^{max}$ obtained using 
(\ref{eq:ssnr}) and actual  $\S_o^{max}$.
\begin{small}
\begin{table}
\begin{center}
\begin{tabular}{|c|c|c|c|c|c|c|}
\hline \hline
$\BW$ & $\sigma_\Noise$ & $\S_i$ &
$\sigma_{\g,opt}$  &  $\sigma_{\g,opt}$&
$\S_{o}^{max}$ & $\S_{o}^{max}$ \\
 &  &  &  (\ref{eq:sopt}) &  & (\ref{eq:ssnr})&  \\
\hline \hline
$8$ & $30$ & $34.9$ & $1.02$ & $0.98$ & $96.3$ & $96.4$ \\ \hline
$8$ & $35$ & $25.7$ & $1.09$ & $1.04$ & $75.7$ & $75.6$ \\ \hline
$8$ & $40$ & $19.7$ & $1.14$ & $1.11$ & $60.9$ & $61.3$ \\ \hline \hline 
$4$ & $35$ & $46.8$ & $0.55$ & $0.57$ & $55.0$ & $71.2$ \\ \hline
$4$ & $40$ & $35.8$ & $0.57$ & $0.60$ & $54.6$ & $57.8$ \\ \hline \hline
$12$ & $30$ & $24.1$ &$ 1.40$ & $1.42$ & $89.3$ & $97.4$ \\ \hline
$12$ & $35$ & $17.7$ & $1.51$ & $1.52$ & $68.0$ & $79.6$ \\ \hline
$12$ & $40$ & $13.6$ & $1.58$ & $1.61$ & $55.0$ & $62.2$ \\ \hline
\end{tabular}
\caption{Comparison of actual $\sigma_\g$, $\S_o^{max}$ with derived
$\sigma_\g$ using (\ref{eq:sopt}), $\S_o^{max}$ using (\ref{eq:ssnr}) for test
signals.}
\label{tab:table2}
\end{center}
\end{table}
\end{small}

\section{Conclusions}
\label{sec:conclusions}

Noise removal is a mandatory pre-processing step in many signal processing
applications. In this paper, we have show that it is possible to identify the
optimal Gaussian filter that best filters noise, under the assumption that the
noise is AWGN. The major contribution of this paper is identification of a
method to obtain the optimal Gaussian filter that best filters a
signal contaminated with AWGN. We have shown experimentally that the identified
method works well for signals whose bandwidth and the input signal to noise ratio 
is know. 
We are in the process of verifying the validity of our approach for 
practical signals like speech.


\bibliographystyle{IEEEtran}
\bibliography{estimating_opt_sigma}

\end{document}